\documentclass[aps, prc, reprint, amsmath, amssymb, superscriptaddress, floatfix, nofootinbib]{revtex4-1} 

\usepackage[pdftex]{graphicx}
\DeclareGraphicsExtensions{.jpg,.png,.pdf}

\usepackage[utf8]{inputenc}
\usepackage{hyperref}
\usepackage{amsmath}
\usepackage{amssymb}
\usepackage{amsfonts}
\usepackage{tabularx}
\usepackage{booktabs}
\usepackage{graphicx}
\usepackage{subfigure}
\usepackage{wrapfig}
\usepackage{float}
\usepackage{color}
\usepackage{multirow}
\usepackage[perpage,symbol]{footmisc}
\usepackage{verbatim}
\usepackage{dcolumn}
\usepackage{bm}
\usepackage[inline]{enumitem}
\definecolor{theblue}{RGB}{0,50,230}
\hypersetup{
  colorlinks=true,
  linkcolor=theblue,
  citecolor=theblue,
  urlcolor=theblue
}

\newcommand{\pt}{\ensuremath{p}_{\rm T}}
\newcommand{\raa}{\ensuremath{R}_{\rm AA}}
\newcommand{\vtwo}{\ensuremath{v}_{\rm 2}}
\newcommand{\snn}{\sqrt{s_{\rm NN}}}

\usepackage{lineno}

\begin{document}

\title{Data-driven extraction of heavy quark diffusion in quark-gluon plasma
}

\author{Shuang~Li}
\email{lish@ctgu.edu.cn}
\affiliation{%
College of Science, China Three Gorges University, Yichang 443002, China\\
}%
\affiliation{Institute of Quantum Matter, South China Normal University, Guangzhou 510006, China.}
\affiliation{%
Key Laboratory of Quark and Lepton Physics (MOE), Central China Normal University, Wuhan 430079, China\\
}%
\affiliation{%
Physics Department and Center for Exploration of Energy and Matter,\\
Indiana University, 2401 N Milo B. Sampson Lane, Bloomington, Indiana 47408, USA
}%
\author{Jinfeng Liao}%
\email{liaoji@indiana.edu}
\affiliation{%
Physics Department and Center for Exploration of Energy and Matter,\\
Indiana University, 2401 N Milo B. Sampson Lane, Bloomington, Indiana 47408, USA
}%

\date{\today}

\begin{abstract}
Heavy quark production provides a unique probe of the quark-gluon plasma transport properties in heavy ion collisions. Experimental observables like the nuclear modification factor $\raa$ and elliptic anisotropy $\vtwo$ of heavy flavor mesons are sensitive to  the heavy quark diffusion coefficient. There now exist an extensive set of such measurements, which allow a data-driven extraction of this coefficient. In this work, we make such an attempt within our recently developed heavy quark transport modeling framework (Langevin-transport with Gluon Radiation, LGR). A question of particular interest is the temperature dependence of the diffusion coefficient, for which we test a wide range of possibility and draw constraints by comparing relevant charm meson data with model results. We find that a relatively strong increase of diffusion coefficient from crossover temperature $T_c$ toward  high temperature is preferred by data. We also make predictions for Bottom meson observables for further experimental tests.  
\end{abstract}


\maketitle

\section{INTRODUCTION}\label{sec:Intro}

At extremely high temperatures such as those available at the earliest moments of cosmic evolution, normal matter turns into a new form of deconfined nuclear matter known as a quark-gluon plasma (QGP). Such a state of matter once filled the early universe when the temperature was high enough. Today the QGP is recreated in laboratories by high energy nuclear collisions at the Relativistic Heavy Ion Collider (RHIC) and the Large Hadron Collider (LHC). A lot of measurements have been performed at RHIC and the LHC, allowing the use of empirical data to extract key properties of the QGP, which are of fundamental interests. The transport properties (such as the shear and bulk viscosity, jet transport coefficient, etc)  have  been found to be particularly informative for  unraveling the dynamical features of QGP, leading to its identification as the nearly perfect fluid~\cite{Gyulassy05,Shuryak05,Muller12}. 

Heavy quark production provides a unique probe of the quark-gluon plasma  in heavy ion collisions~\cite{Rapp:2018qla,Cao:2018ews,Dong:2019byy,Zhou:2014kka,Tang:2014tga,Andronic:2015wma}. Charm and bottom quarks are very hard to be thermally produced in QGP and are dominantly produced from the initial hard scatterings. These rare objects then propagate through the QGP fireball and encode the medium information during their dynamical evolution. Experimental observables  like the nuclear modification factor $\raa$ and elliptic anisotropy $\vtwo$ of (for heavy flavor mesons as well as heavy flavor decay leptons) are sensitive to  the heavy quark diffusion coefficient inside the QGP. There now exist an extensive set of such measurements, which allow a data-driven extraction/constraint of this coefficient. In this work, we make such an attempt within our recently developed heavy quark transport modeling framework, Langevin-transport with Gluon Radiation (LGR)~\cite{Li:2019wri}. 

A particularly interesting question about the QGP transport properties is their temperature dependence, especially how they change in the temperature region from transition temperature $T_c$ to a few times $T_c$. This is the region accessible through RHIC and LHC collision experiments. It has been suggested that such temperature  dependence could be highly nontrivial, especially close to $T_c$. For example, it was proposed long ago that the jet-medium interaction strength (quantified by e.g. normalized jet transport coefficient $\hat{q}/T^3$) may rapidly increase from high temperature down toward $T_c$ and develop a near-$T_c$ peak structure~\cite{Liao:2008dk}. Such a scenario appears to be confirmed by many subsequent studies~\cite{Xu:2014tda,Xu:2015bbz,Burke:2013yra,Shi:2018lsf,Shi:2018vys,Ramamurti:2017zjn}. Another important transport property, shear viscosity over entropy density ratio $\eta/s$, also seems to have a visible T-dependence with a considerable increase from $T_c$ toward higher temperature~\cite{}. Regarding the diffusion and drag coefficients  relevant for heavy quark dynamics, there are also indications of  nontrivial temperature dependence~\cite{}. In this work we will focus on the diffusion coefficient and test a wide range of possibility for its temperature dependence. By comparing modeling results with experimental data of charm hadrons, we draw constraints on the behavior of this important transport property of QGP. Based on that, we further make predictions for bottom hadron observables. 

The rest of this paper is structured as follows. In Section 2, we introduce the detailed setup of the LGR modeling framework and discuss the temperature dependence of diffusion constant. In Section 3, we systematically compare modeling results with data and extract optimal range of this transport coefficient based on global $\chi^2$ analysis. Direct comparison of optimized model results with experimental observables as well as predictions for new measurements are presented in Section 4. Finally we summarize this study in Section 5.

\section{Methodology}\label{sec:Method}

In this Section we present the details of our modeling framework. The heavy quark evolution is described by the following Langevin transport equation that incorporates gluon radiation:~\cite{CaoPRC15}
\begin{equation}
\begin{aligned}\label{eq:LTE_ColRad}
&d\vec{x}=\frac{\vec{p}}{E} dt
& \\
&d\vec{p}=(\vec{F}_{\rm D} + \vec{F}_{\rm T} + \vec{F}_{\rm G}) dt
\end{aligned}
\end{equation}
where the deterministic drag force reads
\begin{equation}\label{eq:DragForce}
\vec{F}_{\rm D}=-\eta_{\rm D}(\vec{p},T) \cdot \vec{p},
\end{equation}
with $\eta_{\rm D}(\vec{p},T)$ being the drag coefficient .
The two-point temporal correlation of the stochastic thermal force $\vec{F}_{T}$  is given by~\cite{RalfSummary16}
\begin{equation}
\begin{aligned}\label{eq:ThermalForceCorre}
&<\vec{F}^{\;\rm i}_{\rm T}(t) \cdot \vec{F}^{\;\rm j}_{\rm T}(t^{\prime})>
= \biggr[ \kappa_{\parallel}(\vec{p},T)P^{\rm ij}_{\parallel} + \kappa_{\perp}(\vec{p},T)P^{\rm ij}_{\perp} \biggr] \delta(t-t^{\prime}),
\end{aligned}
\end{equation}
indicating the uncorrelated random momentum kicks from the medium partons.
$P^{\rm ij}_{\parallel}=p^{i}p^{j}/p^{2}$ and $P^{\rm ij}_{\perp}=\delta^{ij}-p^{i}p^{j}/p^{2}$
are the projection operators for momentum components parallel and perpendicular to the direction of the HQ motion, respectively.
The relation between the drag coefficient ($\eta_{\rm D}$),
the longitudinal ($\kappa_{\parallel}$) and transverse momentum diffusion coefficients ($\kappa_{\perp}$) follows 
\begin{equation}
\begin{aligned}\label{eq:PostPoint1}
\eta_{\rm D}(\vec{p},T)=\frac{\kappa_{\parallel}(\vec{p},T)}{2TE} - \frac{1}{p^{2}} \biggr( \sqrt{\kappa_{\perp}(\vec{p},T)}-\sqrt{\kappa_{\parallel}(\vec{p},T)} \biggr)^{2} \,\, .
\end{aligned}
\end{equation}

The third term on the right hand side of Eq.~\ref{eq:LTE_ColRad},
\begin{equation}\label{eq:RecoilForce}
\vec{F}_{\rm G}=-\sum^{N_{G}}_{j=1}\frac{d\vec{p}^{~\rm j}_{\rm G}}{dt},
\end{equation}
denotes the total recoil force induced by the emitted gluons.
The emission rate of gluons  is estimated with the following Higher-Twist model formula~\cite{HTPRL04}:
\begin{equation}\label{eq:HigherTwist}
\frac{dN_{G}}{dz dk_{\perp}^{2} dt}=\frac{2\alpha_{s}C_{\rm A}P(z) \hat{q}_{\rm q}}{\pi k_{\perp}^{4}}
\biggr[\frac{k_{\perp}^{2}}{k_{\perp}^{2}+(zm_{\rm Q})^{2}}\biggr]^{4} sin^{2}\biggr( \frac{t-t_{\rm 0}}{2\tau_{\rm f}} \biggr).
\end{equation}
In the above, $z$ denotes the fraction of energy carried away by the emitted gluon,
and $P(z)$ represents the quark splitting function;
$\alpha_{s}(k_{\perp})$ is the strong coupling constant of QCD at leading order approximation;
$\tau_{\rm f}=2z(1-z)E/[k_{\perp}^{2}+(zm_{\rm Q})^{2}]$ is the gluon formation time; 
$\hat{q}_{\rm q}$ is the quark jet transport coefficient. The recoil force becomes important when the heavy quark is in the high energy regime $E \gg m_{\rm Q}$, where HQ velocity $v_{Q}=\sqrt{1-(m_{\rm Q}/E)^{2}} \sim 1$ and radiative energy loss becomes significant. In this regime, the $\hat{q}_{\rm q}$ in the above  can be approximated by
\begin{equation}\label{eq:KapaQhat}
\hat{q}_{\rm q}=\frac{2\kappa_{\perp}}{v_{\rm Q}} \approx 2\kappa_{\perp} \, \, .
\end{equation}

As one can see at this point, the key parameters controlling all the forces in Eq.~\ref{eq:LTE_ColRad} are the momentum diffusion coefficients ($\kappa_{\parallel}$ and $\kappa_{\perp}$). It is customary in heavy quark phenomenological modelings  ~\cite{CaoPRC15,AYPRC09,HFModelHee13,CaoPRC15,DukeBayesianPRC17,CTGUHybrid1} to further connect these parameters to the spatial diffusion constant under reasonable approximations. In the low momentum regime where diffusion dynamics is most important, one could assume approximate  isotropy for the momentum diffusion coefficients, i.e.   
$\kappa_{\parallel} = \kappa_{\perp} \equiv \kappa$. Thus the  Eq.~\ref{eq:PostPoint1} is further reduced to
\begin{equation}\label{eq:PostPoint2}
\eta_{\rm D}=\frac{\kappa}{2TE},
\end{equation}
i.e. the so-called dissipation-fluctuation relation in the non-relativistic approximation. 
The connection to the (scaled) spacial diffusion constant, strictly speaking, is valid at zero-momentum limit~\cite{Moore04}, 
$2\pi TD_{s} = 2\pi T^2/[m_{\rm Q} \cdot \eta_{\rm D}(|\vec{p}|\rightarrow0,T)]$. Such relation has been phenomenologically generalized to finite momentum and widely used in heavy quark  modelings~\cite{CaoPRC15,DukeBayesianPRC17,CTGUHybrid1}, allowing the expressions of both the drag and the momentum diffusion coefficients (Eq.~\ref{eq:PostPoint2}) 
in terms of spacial diffusion constant:   
\begin{equation}
\begin{aligned}\label{eq:LTECoef}
&\eta_{\rm D}(\vec{p},T)=\frac{1}{2\pi TD_{s}} \cdot \frac{2\pi T^{2}}{E} \\
&\kappa(T)=\frac{1}{2\pi TD_{s}} \cdot {4\pi T^{3}}.
\end{aligned}
\end{equation} 
We can see that now there is only one key transport parameter, the spatial diffusion constant ($2\pi TD_{s}$) that quantifies all relevant components: the drag force (Eq.~\ref{eq:DragForce}), thermal random force (Eq.~\ref{eq:ThermalForceCorre})
and the recoil force (Eq.~\ref{eq:RecoilForce} and~\ref{eq:HigherTwist})
in the Langevin approach (Eq.~\ref{eq:LTE_ColRad}).
Thus, the dynamical interactions between the heavy quarks  and the QGP medium are conveniently encoded into $2\pi TD_{s}$. We note that, in a naive way, a small/large spatial diffusion corresponds to a short/long mean-free path and thus strong/weak HQ-medium coupling strength. 

Indeed, many past studies have demonstrated sensitivity of experimental observables (such as $\raa$ and $\vtwo$ of heavy flavor mesons) to this key parameter. It appears that, very similar to the situation of jet energy loss, the $\raa$ is mainly controlled by the HQ-medium interaction on average while the $\vtwo$ is strongly influenced by the temperature dependence of the diffusion constant~\cite{HFModelHee13,PHSDPRC16,MCATHQPRC14,GrecoPLB13}. In this study, we aim to investigate such temperature dependence. Let us focus on the temperature range $(1\sim 3) T_c$ and frame the question in a model-independent way. Consider $D_s(T/T_c)$ as an arbitrary function in this range,  it can always be expressed via a series of polynomials (as long as one can include enough terms): $D_s(T/T_c) = d_0 + d_1 (T/T_c) + d_2 (T/T_c)^2 + ...$ without the need of assuming any theoretically-motivated temperature dependence. In principle, with sufficient experimental data and adequate computing power, one could exploit data-driven extraction of all these coefficients term by term. As a first step, we take only the first two terms, i.e. a constant plus a linear dependence, with the following ansatz:  
\begin{equation}\label{eq:Assum}
2\pi TD_{s}(T) \approx \alpha\frac{T}{T_{c}} + \beta \,\, .
\end{equation}
The two dimensionless parameters in Eq.~\ref{eq:Assum},   the slope $\alpha$ and the intercept $\beta$, will be explored in a very wide range without presuming any reasonable value. Our approach is to compute observables for any given $(\alpha, \beta)$ and let the large set of experimental data decide what would be preferred via $\chi^{2}$ analysis. We will then compare so-extracted spatial diffusion constant with various other results~\cite{GrecoPIPNP19}.

Finally we describe a few detailed aspects of the numerical implementation. When solving the  Langevin transport equation (Eq.~\ref{eq:LTE_ColRad}),
we need the space-time evolution of the medium temperature and the fireball velocity field.
It is simulated in terms of a 3+1 dimensional relativistic viscous hydrodynamics based on the HLLE algorithm ( --- see details in ~{\cite{vhlle}}). 
Concerning the hadronization of HQ,
a ``dual" approach, including both fragmentation and heavy-light coalescence mechanisms,
is utilized when the local temperature is below  $T_{c}=165~{\rm GeV}$.
Following our previous work~\cite{CTGUHybrid1,CTGUHybrid2},
the Braaten fragmentation functions~\cite{FragBraaten93} is employed
with the parameter $r=0.1$~\cite{FragFONLLPRL}.
Within the instantaneous coalescence approach,
the momentum distributions of heavy-flavor mesons ($M$) composed of
a heavy quark ($Q$) and a light anti-quark ($\bar{q}$) reads 
\begin{equation}
\begin{aligned}\label{eq:MesonCoal}
\frac{dN_{\rm M}}{d^{3}\vec{p}_{\rm M}}=g_{\rm M}\int d^{6}\xi_{\rm Q} d^{6}\xi_{\rm\bar{q}} f_{\rm Q} f_{\rm\bar{q}}
{\overline W}_{\rm M}^{\rm (n)}(\vec{y}_{\rm M},\vec{k}_{\rm M}) \delta^{(3)}(\sum{\vec{p}})
\end{aligned}
\end{equation}
where, $g_{\rm M}$ is the spin-color degeneracy factor;
$d^{6}\xi_{i}=d^{3}\vec{x}_{i}d^{3}\vec{p}_{i}$ is the phase-space volume for $i=Q,\bar{q}$;
$f_{i}(\vec{x}_{i},\vec{p}_{i})$ denotes the phase-space distributions;
${\overline W}_{\rm M}^{\rm (n)}$ represents the coalescence probability for $Q\bar{q}$ combination
to form the heavy-flavor meson in the $n^{th}$ excited state,
and it is defined as the overlap integral of the Wigner functions
for the meson and $Q\bar{q}$ pair~\cite{NewCoal16}
\begin{equation}\label{eq:InteWigner2}
{\overline W}_{\rm M}^{\rm (n)}(\vec{y}_{\rm M},\vec{k}_{\rm M})=\frac{\lambda^{n}}{n!} e^{-\lambda}, \qquad
\lambda=\frac{1}{2}\biggr(\frac{\vec{y}_{\rm M}^{\;2}}{\sigma_{\rm M}^{2}}+\sigma_{\rm M}^{2}\vec{k}_{\rm M}^{\;2}\biggr).
\end{equation}
where $\vec{y}_{\rm M}=(\vec{x}_{\rm Q}-\vec{x}_{\rm\bar{q}})$ and
$\vec{k}_{\rm M}=(m_{\rm\bar{q}}\vec{p}_{\rm Q}-m_{\rm Q}\vec{p}_{\rm\bar{q}})/(m_{\rm Q}+m_{\rm\bar{q}})$
are the relative coordinate and the relative momentum, respectively, in the center-of-mass frame of $Q\bar{q}$ pair.
Note that the parton Wigner functions are defined through the Gaussian wave-function,
while for heavy-flavor meson, it is quantified by a harmonic oscillator one~\cite{CoalOriginalDover91}.
The width parameter $\sigma_{\rm M}$ is expressed as~\cite{CTGUHybrid1}
\begin{eqnarray}\label{eq:SigM}
\sigma_{\rm M}^{2} = K \frac{(e_{\rm Q}+e_{\rm\bar{q}})(m_{\rm Q}+m_{\rm\bar{q}})^{2}}{e_{\rm Q}m_{\rm\bar{q}}^{2}
+ e_{\rm\bar{q}}m_{\rm Q}^{2}} \langle r_{\rm M}^{2} \rangle
\end{eqnarray}
where, $K=2/3$ ($K=2/5$) for the ground state $n=0$ ($1^{st}$ excited state $n=1$);
$\langle r_{\rm M}^{2} \rangle$ is the mean-square charge radius of a given species of D-meson,
which is predicted by the light-front quark model~\cite{HwangEPJC02};
$e_{i}$ and $m_{i}$ are the charge and mass of a given parton, respectively.
See Ref.~\cite{CTGUHybrid1} for more details.

\section{Constraining Diffusion Constant}\label{sec:Results}

In this Section we focus on constraining diffusion constant using experimental data. 
With any given set of parameters ($\alpha,\beta$) in Eq.~\ref{eq:Assum}, 
we can calculate the corresponding final observable $y$ for the desired species of D-meson.
Then, a $\chi^{2}$ analysis can be performed by comparing the model predictions with   experimental data
\begin{equation}
\begin{aligned}\label{eq:ChiSqu}
&\chi^{2}=\sum_{i=1}^{N} \biggr( \frac{y^{\rm Data}_{i}-y^{\rm Model}_{i}}{\sigma_{i}} \biggr)^{2}.
\end{aligned}
\end{equation}
In the above $\sigma_i$ is the total uncertainty in data points,
including the statistic and systematic components which are added in quadrature.
$n=N-1$ denotes the degree of freedom ($d.o.f$)
when there are $N$ data points used in the comparison.  In this study, we use an extensive set of LHC data: $D^{0}$, $D^{+}$, $D^{\ast +}$ and $D_{s}^{+}$ collected at mid-rapidity ($|y|<0.5$) in the most central ($0-10\%$)
and semi-central ($30-50\%$) Pb--Pb collisions at $\snn=2.76~{\rm TeV}$~\cite{ALICEDesonPbPb2760RAA,ALICEDsPbPb2760RAA}
and $\snn=5.02~{\rm TeV}$~\cite{ALICEDesonPbPb5020RAA},
as well as the $\vtwo$ data in semi-central ($30-50\%$) collisions~\cite{ALICEDesonPbPb2760V2,ALICEDesonPbPb5020V2}.

\begin{figure*}[!htbp]
\centering
\setlength{\abovecaptionskip}{-0.1mm}
\hspace{-2.0em}
\includegraphics[width=.35\textwidth]{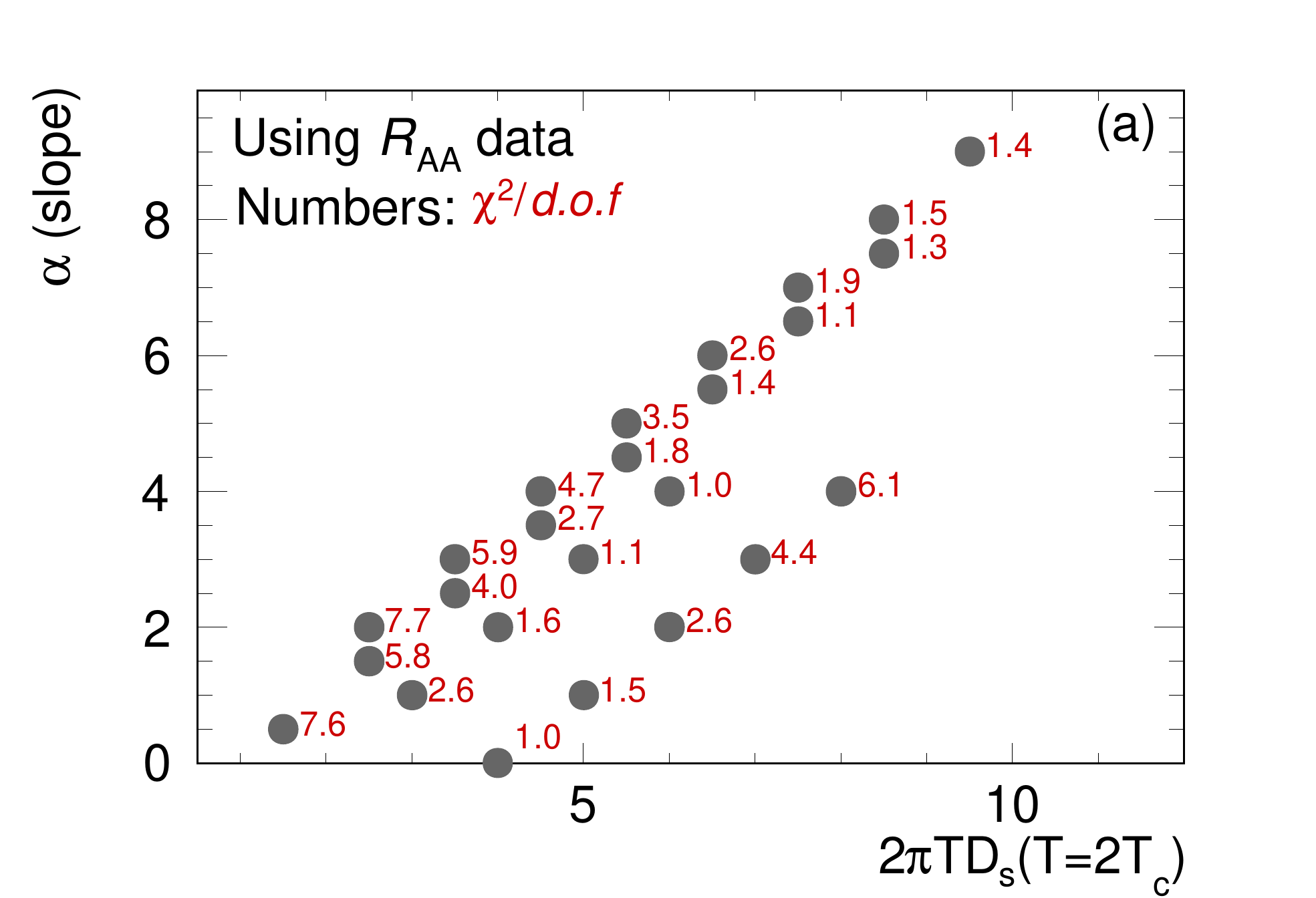}
\hspace{-2.6em}
\includegraphics[width=.35\textwidth]{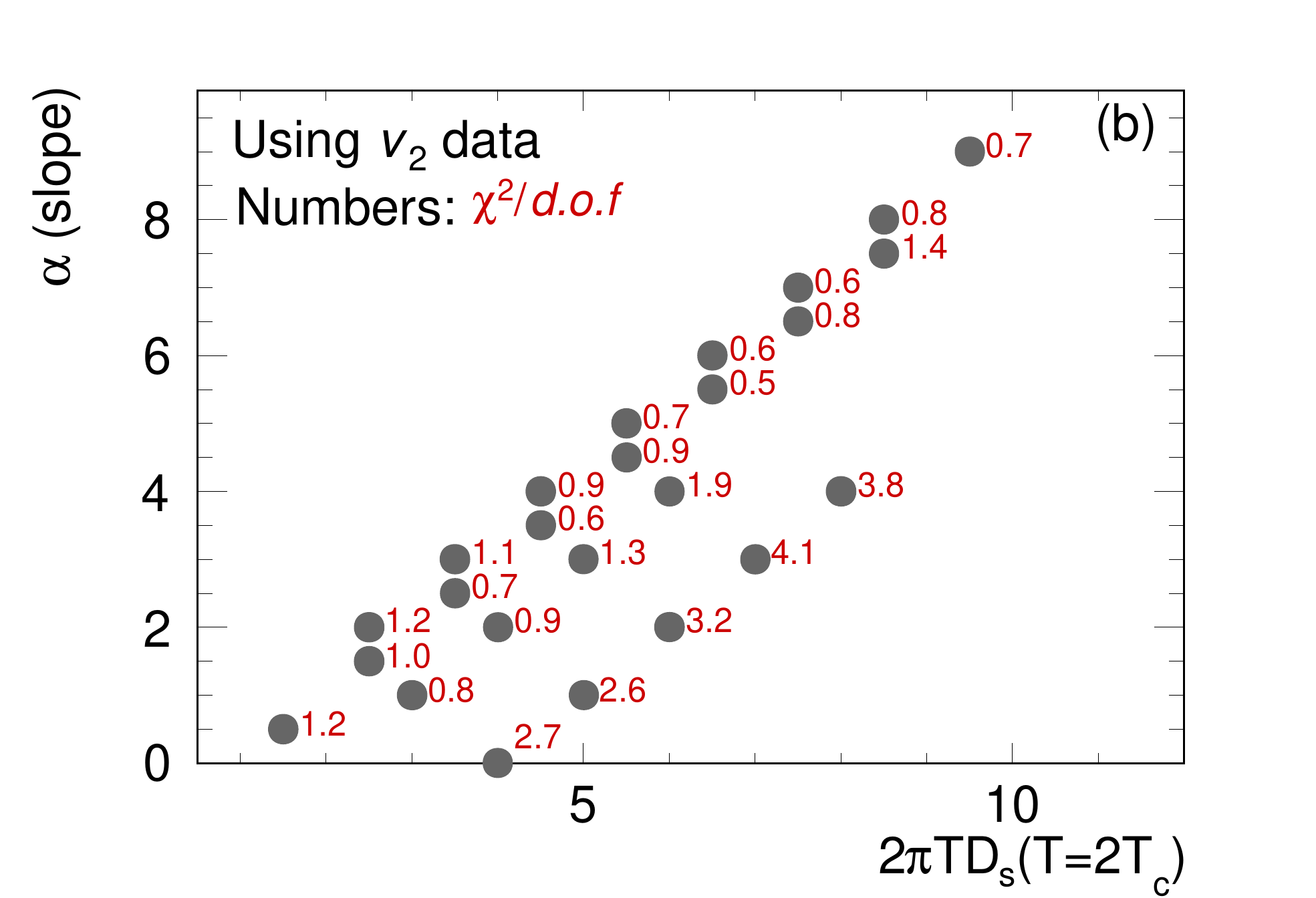}
\hspace{-2.6em}
\includegraphics[width=.35\textwidth]{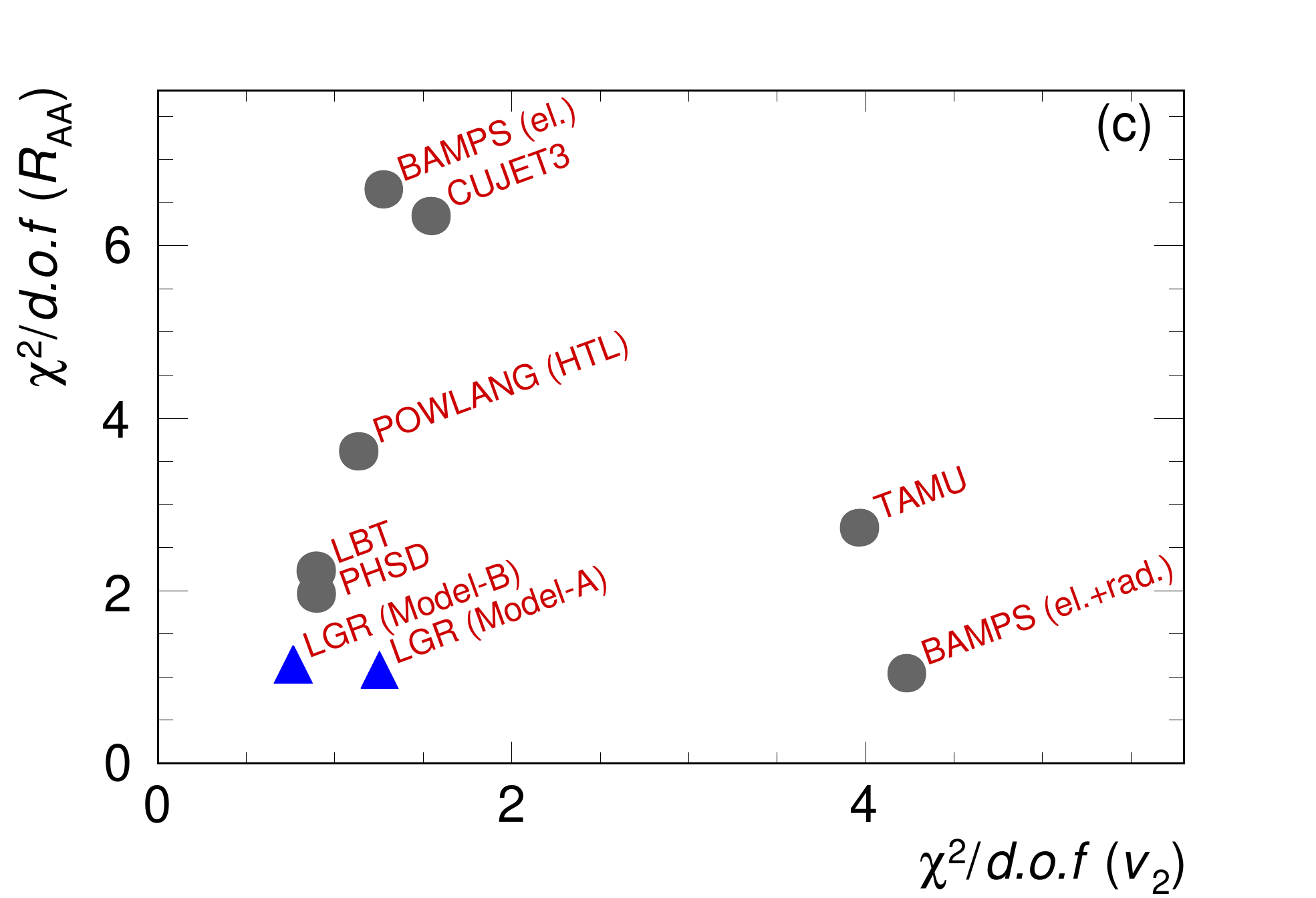}
\hspace{-3.6em}
\caption{(Color online) Comparison of $\chi^{2}/d.o.f$ based on the experimental data of (a) $\raa$ and (b) $\vtwo$.
The model predictions are calculated by using various combinations of parameter $\alpha$ and $\beta$ (Eq.~\ref{eq:Assum}),
which are represented as $\alpha$ (y-axis) and $2\pi TD_{s}(T)$ obtained at $T=2T_{c}$ (x-axis), respectively.
(c) Comparison of $\chi^{2}/d.o.f$ for various models.
See the legend and text for details.}
\label{fig:ChiSquareAllRAAV2DmesonLHC}
\end{figure*}

We scan a wide range of values for $(\alpha,\beta)$ in Eq.~\ref{eq:Assum}: $0\le \alpha\le 9$ and $-8.5\le \beta \le 4$. We note this covers a significantly broader span than existing studies and than commonly conceived reasonable values of $(2\pi T)D_s$. It would be highly unlikely, if not impossible, for the actual QGP diffusion constant to fall outside this range.  A total of 25 different combinations were computed and compared with experimental data, and we summarize these in Tab.~\ref{tab:OptimizedAB}. The $\chi^2$ values were computed separately for $\raa$ and $\vtwo$ as well as for all data combined. To better visualize the results, we also show them in Fig.~\ref{fig:ChiSquareAllRAAV2DmesonLHC}, with panel (a) for  $\raa$ analysis and panel (b) for  $v_{2}$ analysis. In both panels, the y-axis labels the slope $\alpha$ and x-axis labels the $2\pi TD_{s}(T=2T_c)$: basically y-axis quantifies a model's temperature dependence while x-axis calibrates the average diffusion in that model. The different points (filled circles) represent the different combinations of parameters ($\alpha,\beta$) in Tab.~\ref{tab:OptimizedAB}, with the number near each point to display the relevant
$\chi^{2}/d.o.f$ for  that model.   
A number of observations can be drawn from the comprehensive model-data comparison. For the $\raa$, several models achieve $\chi^2/d.o.f \sim 1$ with widespread values of slope parameter. This suggests that $\raa$ appears to be more sensitive to the average diffusion constant while insensitive to the temperature dependence.  For the $\vtwo$, it clearly shows a stronger sensitivity to the temperature dependence. There also exist several models with $\chi^2/d.o.f \sim 1$ and it appears that a small value of $(2\pi T)D_s$ near $T_c$ is crucial for a better description of $\vtwo$ data. Taken all together, we are able to identify two particular models that outperform others in describing both $\raa$ and $\vtwo$ data simultaneously with $\chi^2/d.o.f \sim 1$. These two will be the parameter-optimized models from the LGR framework: the {\bf LGR (Model-A)} with $(\alpha,\beta)= (3,-1)$ and the {\bf LGR (Model-B) with  $(\alpha,\beta)= (6.5,-5.5) $}. While both models  give similarly nice good description of $\raa$, the Model-B has a much stronger temperature dependence and gives a better description of $\vtwo$.

\begin{table}[!htbp]
\centering
\begin{tabular}{c|c|c|c|c|c}
\hline
\hline
\multicolumn{1}{c}{\multirow{2}{*}{$\it Model~ID$}}
 & \multicolumn{1}{c}{\multirow{1}{*}{\centering $\alpha$}}
 & \multicolumn{1}{c}{\multirow{1}{*}{\centering $\beta$}}
 & \multicolumn{1}{c}{\multirow{1}{*}{\centering $\chi^{2}/d.o.f$}}
 & \multicolumn{1}{c}{\multirow{1}{*}{\centering $\chi^{2}/d.o.f$}}
 & \multicolumn{1}{c}{\multirow{2}{*}{\centering Total}}
  \\
\multicolumn{1}{c}{\centering }                                                                                                                              
 & \multicolumn{1}{c}{\centering (Slope)}
 & \multicolumn{1}{c}{\centering (Intercept)}
 & \multicolumn{1}{c}{\centering ($\raa$)}
 & \multicolumn{1}{c}{\centering ($\vtwo$)}
 & \multicolumn{1}{c}{\centering }
 \\
\cline{1-6}
\multicolumn{1}{c}{\centering 1}
 & \multicolumn{1}{c}{\centering 1.00}
 & \multicolumn{1}{c}{\centering 1.00}
 & \multicolumn{1}{c}{\centering 2.61}
 & \multicolumn{1}{c}{\centering 0.85}
 & \multicolumn{1}{c}{\centering 2.37}
 \\
\cline{1-6}
\multicolumn{1}{c}{\centering 2}
 & \multicolumn{1}{c}{\centering 2.00}
 & \multicolumn{1}{c}{\centering 0.00}
 & \multicolumn{1}{c}{\centering 1.58}
 & \multicolumn{1}{c}{\centering 0.87}
 & \multicolumn{1}{c}{\centering 1.49}
 \\
\cline{1-6}
\multicolumn{1}{c}{\centering \textbf{3 (A)}}
 & \multicolumn{1}{c}{\centering \textbf{3.00}}
 & \multicolumn{1}{c}{\centering \textbf{-1.00}}
 & \multicolumn{1}{c}{\centering \textbf{1.09}}
 & \multicolumn{1}{c}{\centering \textbf{1.26}}
 & \multicolumn{1}{c}{\centering \textbf{1.11}}
 \\
\cline{1-6}
\multicolumn{1}{c}{\centering 4}
 & \multicolumn{1}{c}{\centering 4.00}
 & \multicolumn{1}{c}{\centering -2.00}
 & \multicolumn{1}{c}{\centering 1.03}
 & \multicolumn{1}{c}{\centering 1.86}
 & \multicolumn{1}{c}{\centering 1.14}
 \\
\cline{1-6}
\multicolumn{1}{c}{\centering 5}
 & \multicolumn{1}{c}{\centering 0.00}
 & \multicolumn{1}{c}{\centering 4.00}
 & \multicolumn{1}{c}{\centering 1.03}
 & \multicolumn{1}{c}{\centering 2.68}
 & \multicolumn{1}{c}{\centering 1.26}
 \\
\cline{1-6}
\multicolumn{1}{c}{\centering 6}
 & \multicolumn{1}{c}{\centering 1.00}
 & \multicolumn{1}{c}{\centering 3.00}
 & \multicolumn{1}{c}{\centering 1.55}
 & \multicolumn{1}{c}{\centering 2.64}
 & \multicolumn{1}{c}{\centering 1.70}
 \\
\cline{1-6}
\multicolumn{1}{c}{\centering 7}
 & \multicolumn{1}{c}{\centering 2.00}
 & \multicolumn{1}{c}{\centering 2.00}
 & \multicolumn{1}{c}{\centering 2.58}
 & \multicolumn{1}{c}{\centering 3.17}
 & \multicolumn{1}{c}{\centering 2.66}
 \\
\cline{1-6}
\multicolumn{1}{c}{\centering 8}
 & \multicolumn{1}{c}{\centering 3.00}
 & \multicolumn{1}{c}{\centering 1.00}
 & \multicolumn{1}{c}{\centering 4.45}
 & \multicolumn{1}{c}{\centering 4.06}
 & \multicolumn{1}{c}{\centering 4.40}
 \\
\cline{1-6}
\multicolumn{1}{c}{\centering 9}
 & \multicolumn{1}{c}{\centering 4.00}
 & \multicolumn{1}{c}{\centering 0.00}
 & \multicolumn{1}{c}{\centering 6.10}
 & \multicolumn{1}{c}{\centering 3.76}
 & \multicolumn{1}{c}{\centering 5.78}
 \\
\cline{1-6}
\multicolumn{1}{c}{\centering 10}
 & \multicolumn{1}{c}{\centering 0.50}
 & \multicolumn{1}{c}{\centering 0.50}
 & \multicolumn{1}{c}{\centering 7.61}
 & \multicolumn{1}{c}{\centering 1.17}
 & \multicolumn{1}{c}{\centering 6.73}
 \\
\cline{1-6}
\multicolumn{1}{c}{\centering 11}
 & \multicolumn{1}{c}{\centering 1.50}
 & \multicolumn{1}{c}{\centering -0.50}
 & \multicolumn{1}{c}{\centering 5.77}
 & \multicolumn{1}{c}{\centering 1.04}
 & \multicolumn{1}{c}{\centering 5.12}
 \\
\cline{1-6}
\multicolumn{1}{c}{\centering 12}
 & \multicolumn{1}{c}{\centering 2.50}
 & \multicolumn{1}{c}{\centering -1.50}
 & \multicolumn{1}{c}{\centering 3.95}
 & \multicolumn{1}{c}{\centering 0.70}
 & \multicolumn{1}{c}{\centering 3.51}
 \\
\cline{1-6}
\multicolumn{1}{c}{\centering 13}
 & \multicolumn{1}{c}{\centering 3.50}
 & \multicolumn{1}{c}{\centering -2.50}
 & \multicolumn{1}{c}{\centering 2.68}
 & \multicolumn{1}{c}{\centering 0.58}
 & \multicolumn{1}{c}{\centering 2.39}
 \\
\cline{1-6}
\multicolumn{1}{c}{\centering 14}
 & \multicolumn{1}{c}{\centering 4.50}
 & \multicolumn{1}{c}{\centering -3.50}
 & \multicolumn{1}{c}{\centering 1.84}
 & \multicolumn{1}{c}{\centering 0.90}
 & \multicolumn{1}{c}{\centering 1.71}
 \\
\cline{1-6}
\multicolumn{1}{c}{\centering 15}
 & \multicolumn{1}{c}{\centering 5.50}
 & \multicolumn{1}{c}{\centering -4.50}
 & \multicolumn{1}{c}{\centering 1.44}
 & \multicolumn{1}{c}{\centering 0.53}
 & \multicolumn{1}{c}{\centering 1.32}
 \\
\cline{1-6}
\multicolumn{1}{c}{\centering \textbf{16 (B)}}
 & \multicolumn{1}{c}{\centering \textbf{6.50}}
 & \multicolumn{1}{c}{\centering \textbf{-5.50}}
 & \multicolumn{1}{c}{\centering \textbf{1.15}}
 & \multicolumn{1}{c}{\centering \textbf{0.77}}
 & \multicolumn{1}{c}{\centering \textbf{1.10}}
\\
\cline{1-6}
\multicolumn{1}{c}{\centering 17}
 & \multicolumn{1}{c}{\centering 7.50}
 & \multicolumn{1}{c}{\centering -6.50}
 & \multicolumn{1}{c}{\centering 1.27}
 & \multicolumn{1}{c}{\centering 1.41}
 & \multicolumn{1}{c}{\centering 1.29}
 \\
\cline{1-6}
\multicolumn{1}{c}{\centering 18}
 & \multicolumn{1}{c}{\centering 2.00}
 & \multicolumn{1}{c}{\centering -1.50}
 & \multicolumn{1}{c}{\centering 7.69}
 & \multicolumn{1}{c}{\centering 1.25}
 & \multicolumn{1}{c}{\centering 6.81}
 \\
\cline{1-6}
\multicolumn{1}{c}{\centering 19}
 & \multicolumn{1}{c}{\centering 3.00}
 & \multicolumn{1}{c}{\centering -2.50}
 & \multicolumn{1}{c}{\centering 5.95}
 & \multicolumn{1}{c}{\centering 1.05}
 & \multicolumn{1}{c}{\centering 5.28}
 \\
\cline{1-6}
\multicolumn{1}{c}{\centering 20}
 & \multicolumn{1}{c}{\centering 4.00}
 & \multicolumn{1}{c}{\centering -3.50}
 & \multicolumn{1}{c}{\centering 4.75}
 & \multicolumn{1}{c}{\centering 0.89}
 & \multicolumn{1}{c}{\centering 4.22}
 \\
\cline{1-6}
\multicolumn{1}{c}{\centering 21}
 & \multicolumn{1}{c}{\centering 5.00}
 & \multicolumn{1}{c}{\centering -4.50}
 & \multicolumn{1}{c}{\centering 3.46}
 & \multicolumn{1}{c}{\centering 0.68}
 & \multicolumn{1}{c}{\centering 3.08}
 \\
\cline{1-6}
\multicolumn{1}{c}{\centering 22}
 & \multicolumn{1}{c}{\centering 6.00}
 & \multicolumn{1}{c}{\centering -5.50}
 & \multicolumn{1}{c}{\centering 2.61}
 & \multicolumn{1}{c}{\centering 0.60}
 & \multicolumn{1}{c}{\centering 2.34}
 \\
\cline{1-6}
\multicolumn{1}{c}{\centering 23}
 & \multicolumn{1}{c}{\centering 7.00}
 & \multicolumn{1}{c}{\centering -6.50}
 & \multicolumn{1}{c}{\centering 1.93}
 & \multicolumn{1}{c}{\centering 0.57}
 & \multicolumn{1}{c}{\centering 1.74}
 \\
\cline{1-6}
\multicolumn{1}{c}{\centering 24}
 & \multicolumn{1}{c}{\centering 8.00}
 & \multicolumn{1}{c}{\centering -7.50}
 & \multicolumn{1}{c}{\centering 1.52}
 & \multicolumn{1}{c}{\centering 0.76}
 & \multicolumn{1}{c}{\centering 1.42}
 \\
\cline{1-6}
\multicolumn{1}{c}{\centering 25}
 & \multicolumn{1}{c}{\centering 9.00}
 & \multicolumn{1}{c}{\centering -8.50}
 & \multicolumn{1}{c}{\centering 1.35}
 & \multicolumn{1}{c}{\centering 0.66}
 & \multicolumn{1}{c}{\centering 1.26}
 \\
\hline
\hline
\end{tabular}
\caption{Summary of the adjustable parameters in Eq.~\ref{eq:Assum},
together with the relevant $\chi^{2}/d.o.f$ obtained for $\raa$ and $\vtwo$.}
\label{tab:OptimizedAB}
\end{table}

Let us make a comparison with various existing modeling frameworks, e.g. TAMU~\cite{HFModelHee13}, PHSD~\cite{PHSDPRC16},
LTB~\cite{LBTPRC16}, POWLANG~\cite{POWLANGEPJC11},
BAMPS (eastic)~\cite{BAMPS10}, BAMPS (elastic+radiative)~\cite{BAMPS10} and CUJET3~\cite{CUJET3CPC18,CUJET3Arxiv18}.  
The published results from these models for relevant observables are taken from pertinent references and used to evaluate the corresponding  $\chi^{2}$ for each model. The analysis results $\chi^{2}/d.o.f(\raa)$ and $\chi^{2}/d.o.f(\vtwo)$ are then shown and compared in the panel (c) of Fig.~\ref{fig:ChiSquareAllRAAV2DmesonLHC}.  
One can see that to describe simultaneously both  $\raa$ and $\vtwo$ data is challenging in general.  The LGR Model-A and Model-B, featuring a moderate to strong temperature dependence and a small diffusion constant very close to $T_c$, demonstrate a successful description of current measurements.

\begin{figure}[!htbp]
\centering
\setlength{\abovecaptionskip}{-0.1mm}
\includegraphics[width=.4\textwidth]{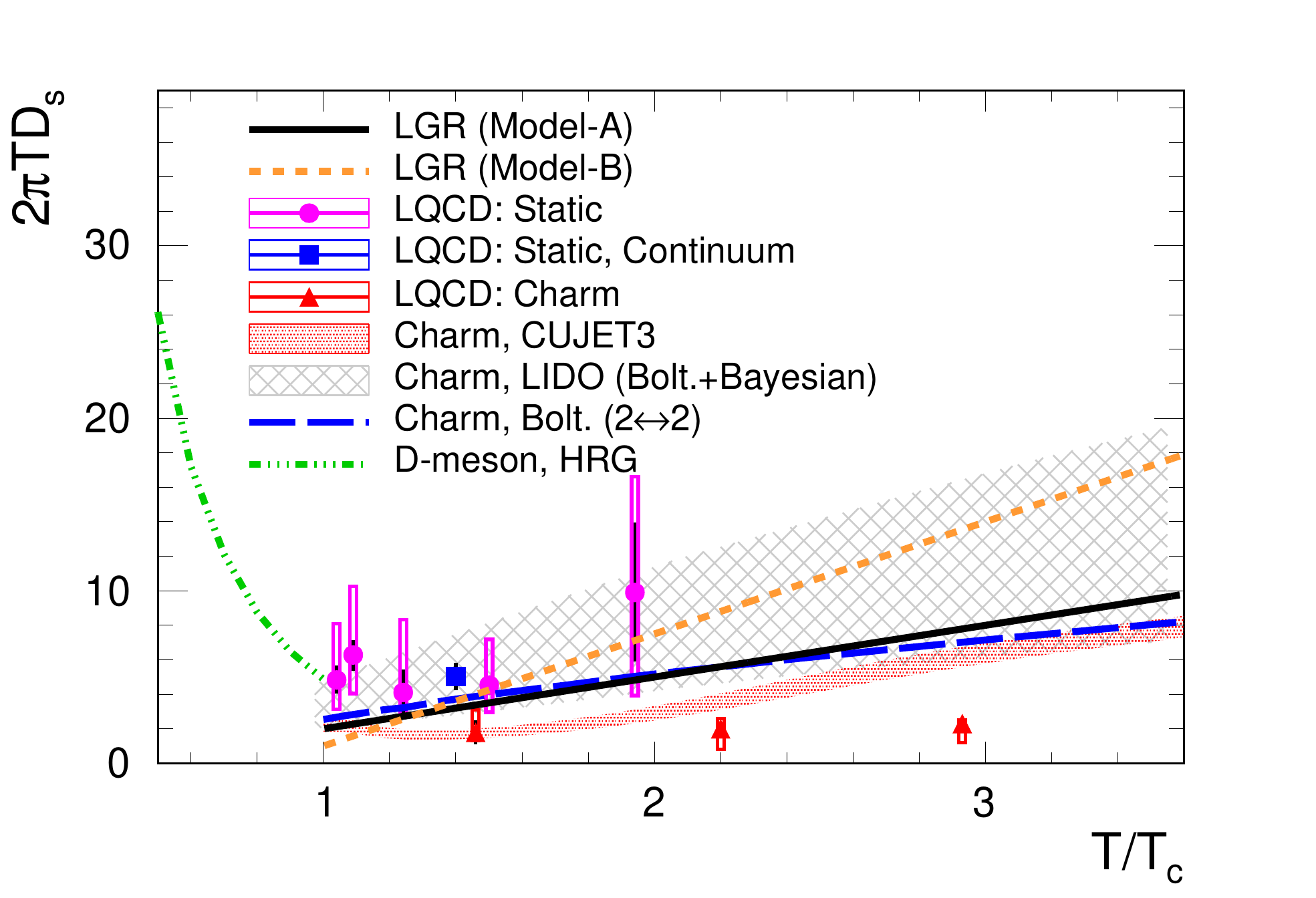}
\caption{(Color online) Spatial diffusion constant $2\pi TD_{s}(T)$ of charm quark
from various calculations, including: the LGR Model-A and Model-B, the optimized parameters, lattice QCD calculations (pink circle~\cite{LQCDbanerjee12}, red triangle~\cite{LQCDding12}
and blue square~\cite{LQCDolaf14}), CUJET3(red region~\cite{CUJET3JHEP16}),
a Bayesian analysis  in $95\%~CL$ from LIDO (shadowed gray band~\cite{Lido18})
and a LO calculation with a Boltzmann dynamics (long dashed blue curve~\cite{CTGUHybrid3}). 
The result for D-meson (dot dashed green curve~\cite{2PiTDs4Dmeson}) in the hadronic phase is also shown for comparison.}
\label{fig:OptimizedCoef}
\end{figure}

Finally in Figure~\ref{fig:OptimizedCoef} we present the spacial diffusion constant $2\pi TD_{s}$ of charm quark from various phenomenological extractions and theoretical calculations. Very close to $T_c$, the results from both LGR Model-A (solid black curve) and Model-B (dashed orange curve)   are compatible with the LQCD calculations within their
significant uncertainties (pink circle~\cite{LQCDbanerjee12}, red triangle~\cite{LQCDding12}
and blue square~\cite{LQCDolaf14}) as well as consistent with other models~\cite{CTGUHybrid3,ADSCFTOleg, GubserPRD07}.  
Toward the higher temperature end, the region spanned by our Model-A and Model-B compare well with the band from  the Bayesian analysis based on the Duke model (gray region~\cite{Lido18}). Combing various information together, we observe that: (1) a small value $2\pi TD_{s} \simeq (2\sim 4)$ appears to be much preferred in the vicinity of $T_c$; (2) a relatively strong increase of its value toward higher temperature is favored, albeit still with large uncertainty for $T\gtrsim 2T_c$.

\section{LGR Results for Observables}\label{sec:observables}

In this Section, we present the results for various observables to be compared with experimental data. Specifically we use the optimized LGR Model-B based on analysis from the previous Section.

\begin{figure}[!htbp]
    \centering
    \includegraphics[width=0.8\linewidth]{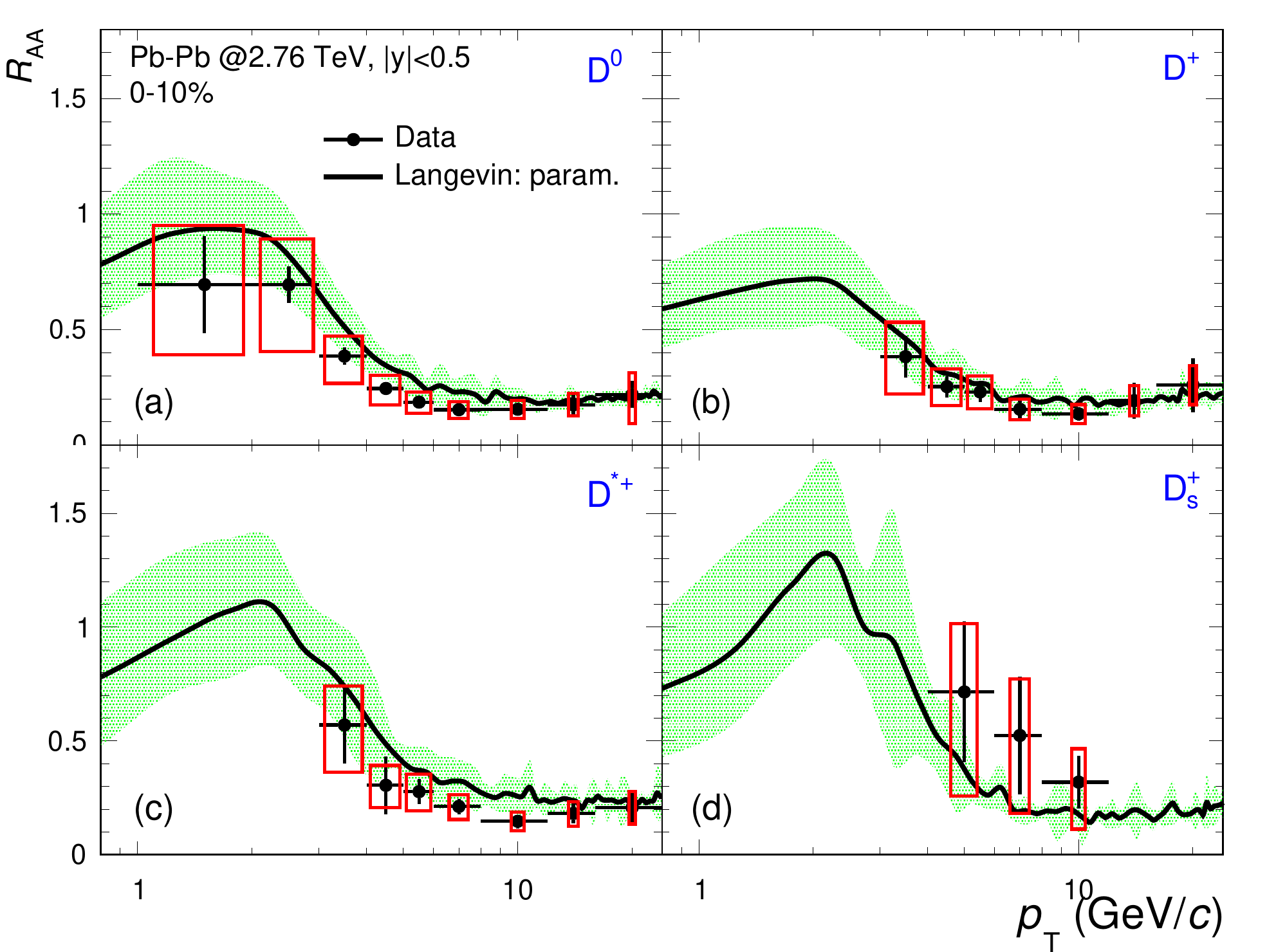}
    \caption{(Color online) Comparison between experimental data (red box~\cite{ALICEDesonPbPb2760RAA, ALICEDsPbPb2760RAA}) and LGR Model-B calculations (solid black curve with green uncertainty band) for the nuclear modification factor $\raa$, of (a) $D^{0}$, (b) $D^{+}$, (c) $D^{\ast +}$ and (d) $D_{s}^{+}$ at mid-rapidity ($|y|<0.5$) in the  ($0-10\%$) centrality Pb--Pb collisions at $\snn=2.76~{\rm TeV}$. }
    \label{fig:RAADmeson2760}
 \end{figure}

\begin{figure}[!htbp]
    \centering
    \includegraphics[width=0.8\linewidth]{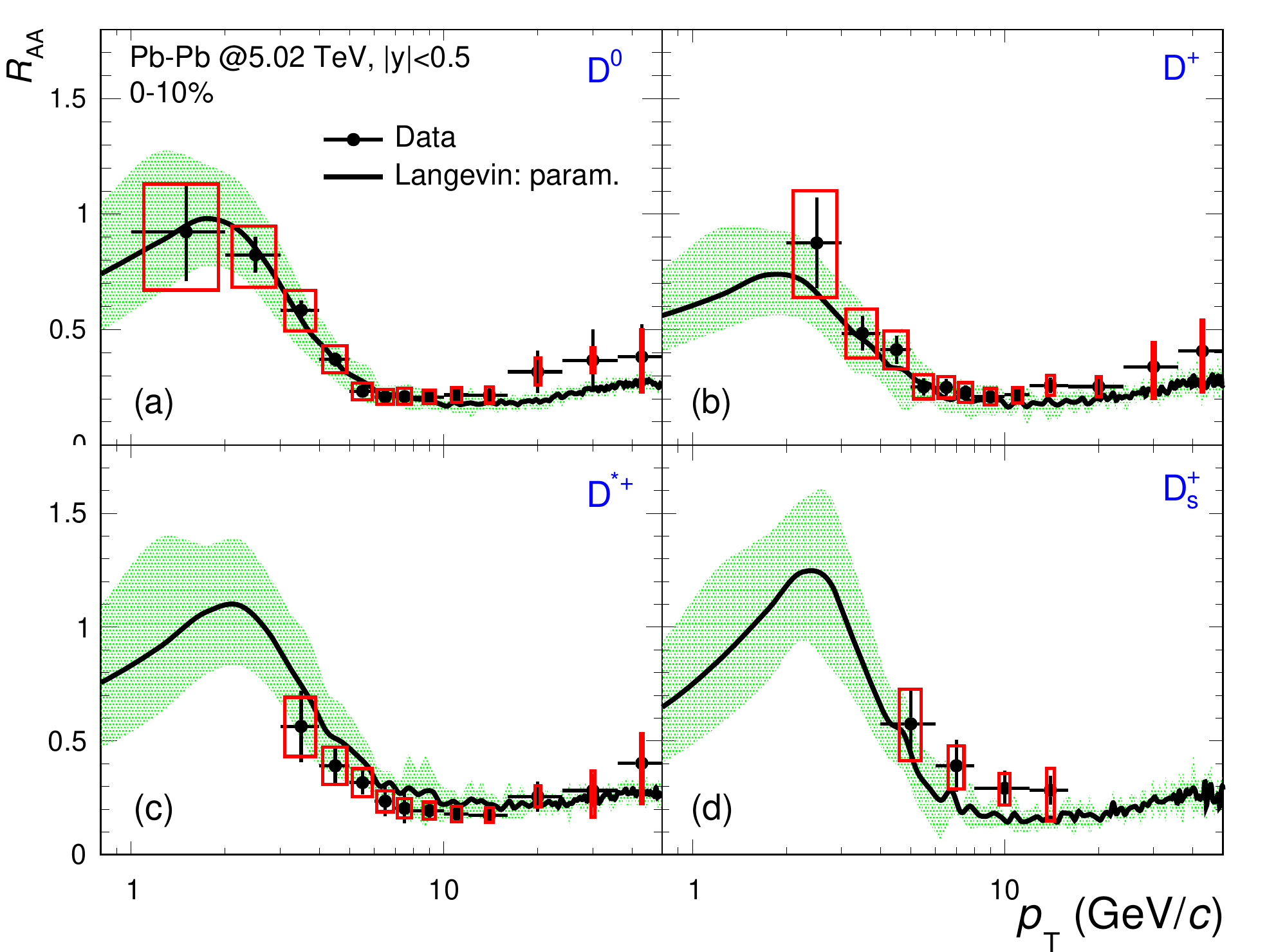}
    \caption{Comparison between experimental data (red box~\cite{ALICEDesonPbPb5020RAA}) and LGR Model-B calculations (solid black curve with green uncertainty band) for the nuclear modification factor $\raa$, of (a) $D^{0}$, (b) $D^{+}$, (c) $D^{\ast +}$ and (d) $D_{s}^{+}$ at mid-rapidity ($|y|<0.5$) in the  ($0-10\%$) centrality Pb--Pb collisions at $\snn=5.02~{\rm TeV}$.}
    \label{fig:RAADmeson5020}
 \end{figure}

Figure~\ref{fig:RAADmeson2760} shows the $\raa$
of (a) $D^{0}$, (b) $D^{+}$, (c) $D^{\ast +}$ and (d) $D_{s}^{+}$
in the most central ($0-10\%$) Pb--Pb collisions at $\snn=2.76~{\rm TeV}$, respectively. 
The calculations are done with FONLL  initial charm quark spectra and EPS09 NLO  parametrization for the nPDF in Pb~\cite{CTGUHybrid1}, and the green band reflects the theoretical uncertainties coming from these inputs. 
It can be seen that   the model calculations 
provide a very good description of the measured $\pt$-dependent $\raa$ data for various charm mesons.  The same conclusion can be drawn for the comparison in Pb--Pb collisions at $\snn=5.02~{\rm TeV}$, as shown in Fig.~\ref{fig:RAADmeson5020}.

\begin{figure}[!btbp]
\centering
\setlength{\abovecaptionskip}{-0.1mm}
\includegraphics[width=.4\textwidth]{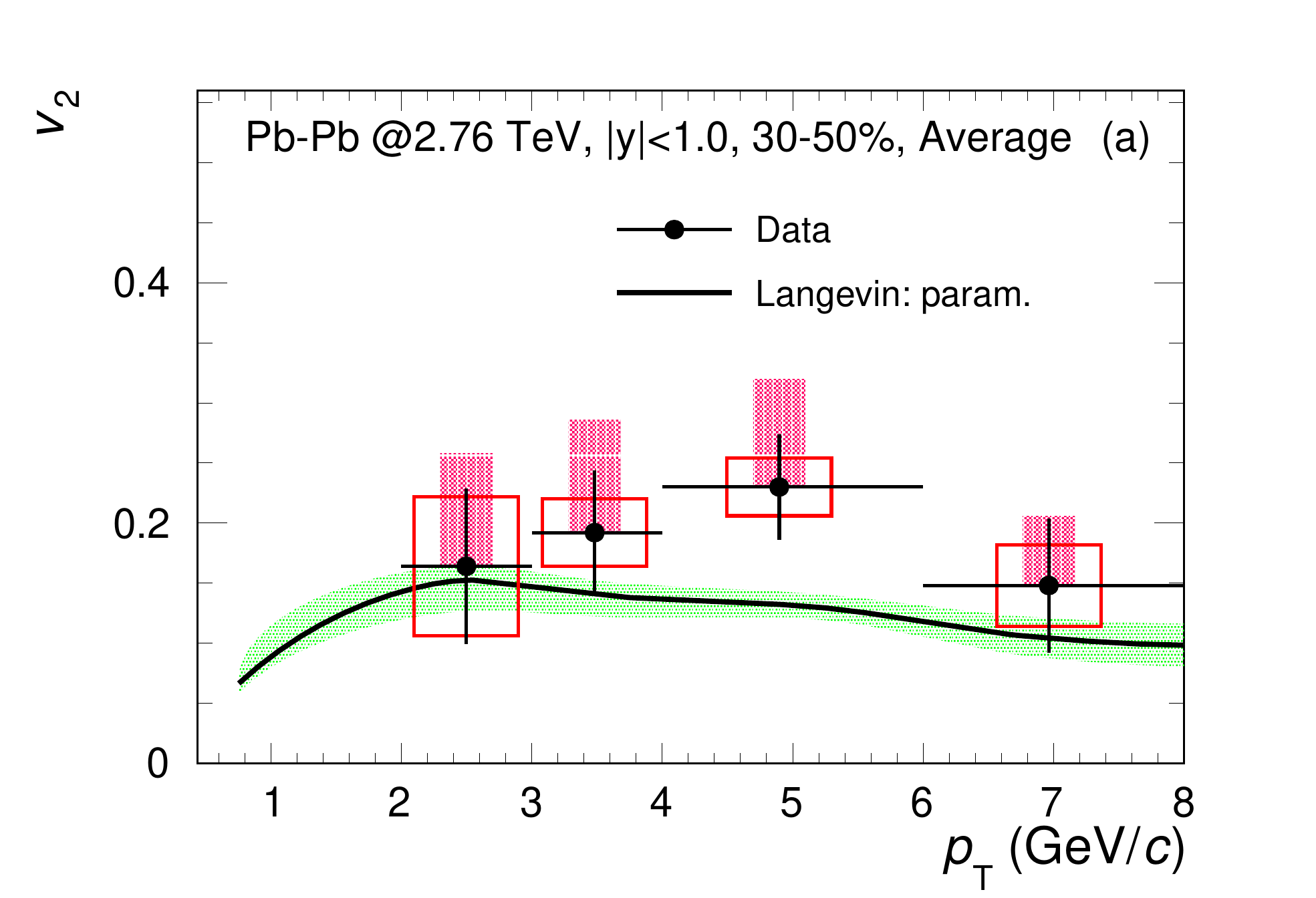}
\includegraphics[width=.4\textwidth]{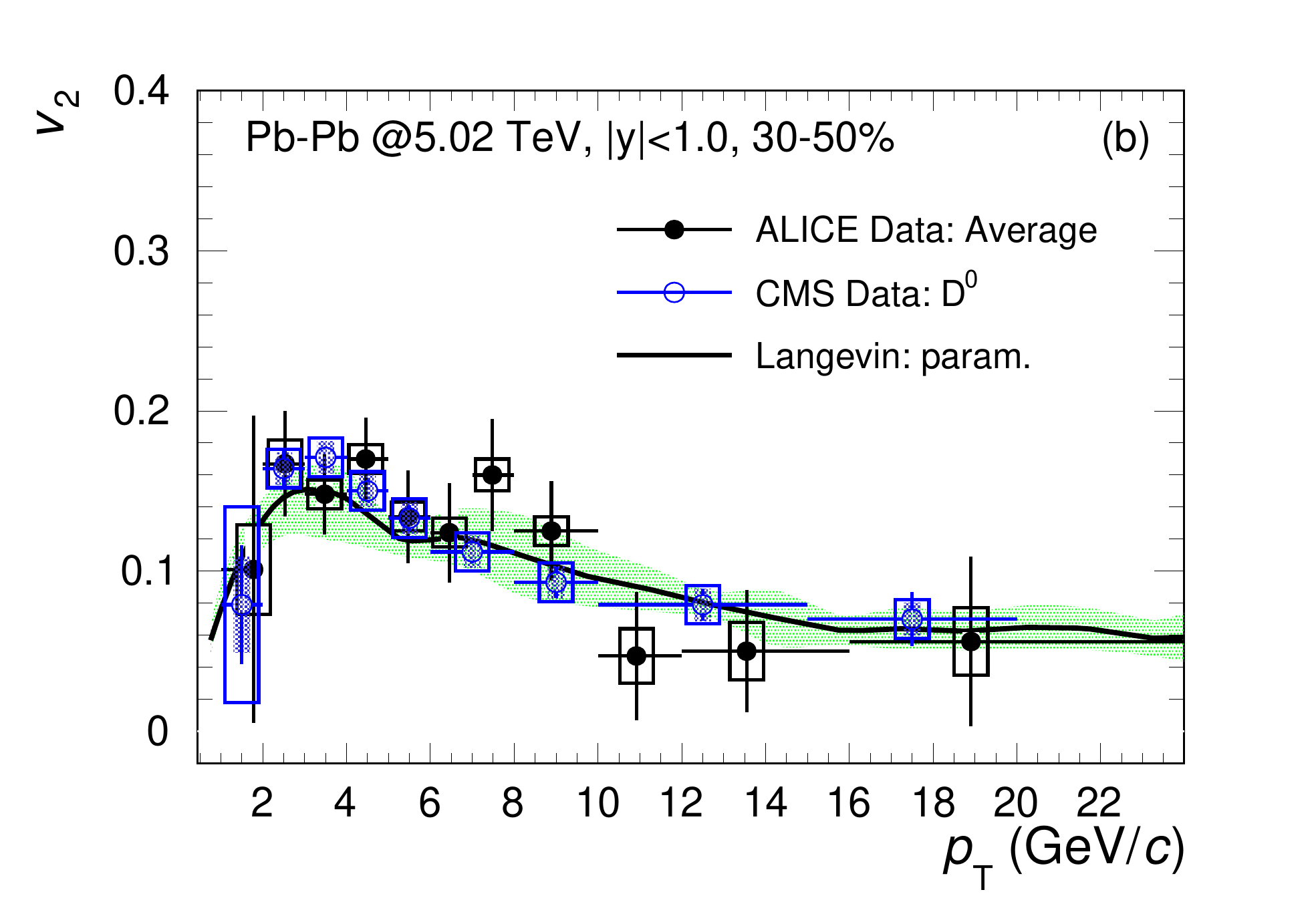}
\caption{(Color online) 
Comparison between experimental data (red~\cite{ALICEDesonPbPb2760V2}, black~\cite{ALICEDesonPbPb5020V2} and blue boxes~\cite{CMSD0PbPb5020V2}) and LGR Model-B calculations (solid black curve with green uncertainty band) 
for the elliptic flow $\vtwo$ of non-strange D-meson at mid-rapidity ($|y|<0.5$)
in the $30-50\%$ centrality Pb--Pb collisions at (a) $\snn=2.76~{\rm TeV}$ and (b) $\snn=5.02~{\rm TeV}$. }
\label{fig:V2Dmeson2760_5020_C1}
\end{figure}

Figure~\ref{fig:V2Dmeson2760_5020_C1} presents the elliptic flow coefficient $\vtwo$ of
non-strange D-meson (averaged $D^{0}$, $D^{+}$, and $D^{\ast +}$)
in the $30-50\%$ centrality Pb--Pb collisions at (a) $\snn=2.76~{\rm TeV}$ and (b) $\snn=5.02~{\rm TeV}$.
Within the uncertainties of the experimental data,
our model calculations describe well the anisotropy of the
transverse momentum distribution of the non-strange D-meson.
The sizable $\vtwo$ of these charm mesons, in particular at intermediate $\pt\sim3-5~{\rm GeV}$,
suggests that charm quarks actively participate in the collective expansion of the fireball.

\begin{figure}[!htbp]
\centering
\setlength{\abovecaptionskip}{-0.1mm}
\includegraphics[width=.4\textwidth]{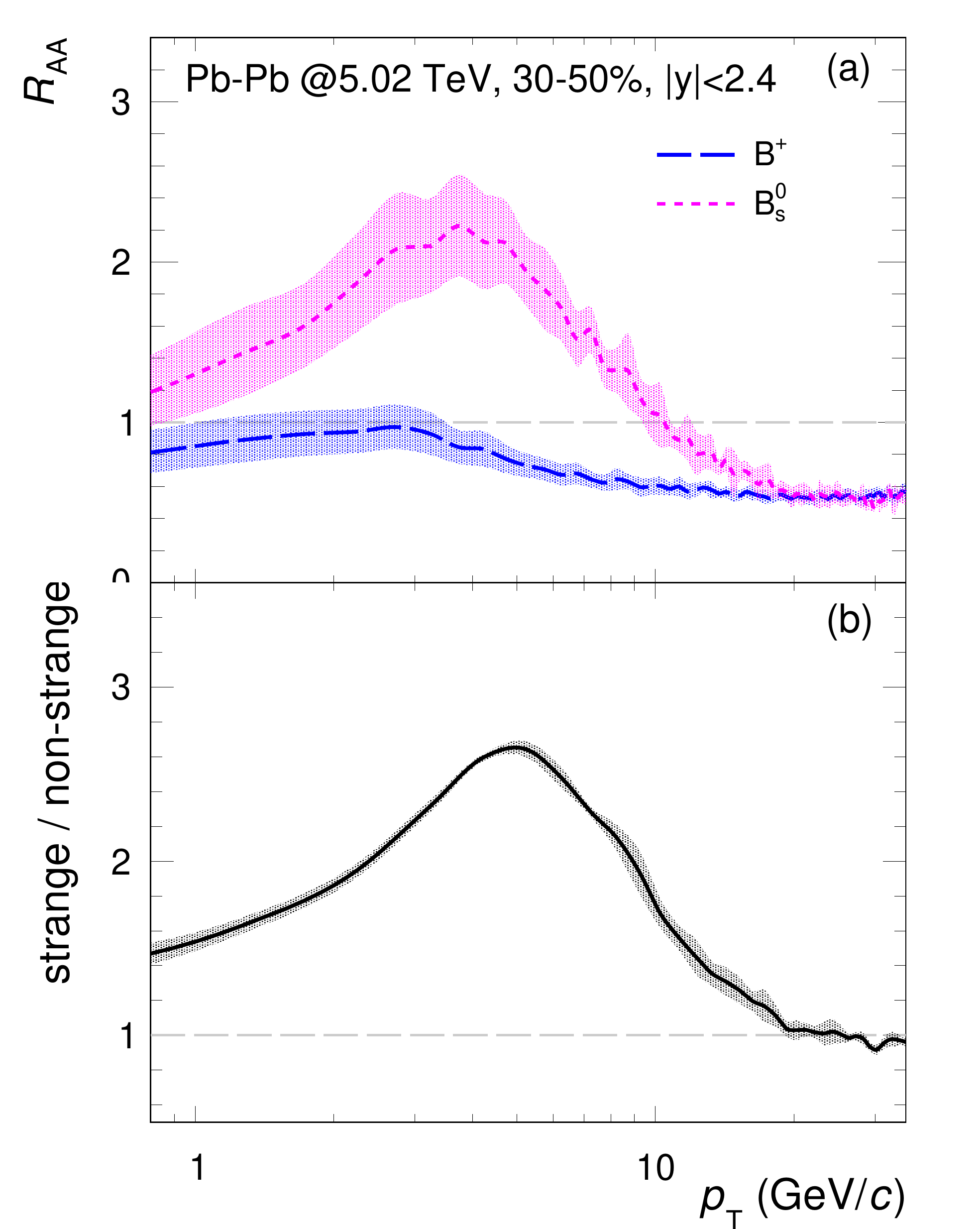}
\caption{(Color online) (a) Comparison of $\raa(B^{+})$ (solid black curve) and $\raa(B_{s}^{0})$ (dashed red curve)
in semi-central ($30-50\%$) Pb--Pb collisions at $\snn=5.02~{\rm TeV}$.
(b) $\raa$ ratio between $\raa(B_{s}^{0})$ and $\raa(B^{+})$.}
\label{fig:Bottom}
\end{figure}

Given that our model has provided a very good description of charm meson data, it is tempting to further test it with bottom meson measurements. Here we present LGR Model-B results   for the strange and non-strange bottom mesons. To do that, one would need the relevant transport coefficient for bottom quarks. It has been suggested~\cite{CTGUHybrid3,POWLANGEPJC11,DasPRD16} that the ratio of bottom quark spacial diffusion
constant  to that for charm quark exhibits a weak $T$-dependence and varies within $\sim0.8-0.9$ in the range $T_{c}<T<4T_{c}$. We therefore use a constant factor $0.85$ to give a temperature dependent spatial diffusion constant $2\pi TD_{s}(bottom) = 0.85\times 2\pi TD_{s}(charm)$ for calculating  the nuclear modification factor
of open-bottom hadrons.
Figure~\ref{fig:Bottom} shows the obtained results in the $30-50\%$ centrality 
Pb--Pb collisions at $\snn=5.02~{\rm TeV}$. 
It is found that $\raa(B_{s}^{0})$ is significant larger than $\raa(B^{+})$,
in particular at $\pt\sim4-6~{\rm GeV}$. This difference decreases  toward high $\pt$. 
Similar to previous results for the open-charm systems~\cite{CTGUHybrid2}, the enhancement behavior is mainly induced by the heavy-light coalescence effect, which is more pronounced for the $B_{s}^{0}(\bar{b}s)$
than for the $B^{+}(\bar{b}u)$. 
The observation is consistent with the B-meson measurements ($0-100\%$) reported by the CMS Collaboration~\cite{BMesonCMSPLB19}.

\section{Summary}\label{sec:Summary}

In summary we have used a recently developed heavy quark transport modeling framework (Langevin-transport with Gluon Radiation, LGR) to study the heavy flavor spatial diffusion constant in the quark-gluon plasma in a data-driven approach. In particular we've examined the temperature dependence of this transport coefficient by systematically scanning a wide range of possibilities. Our global $\chi^2$ analysis using extensive set of LHC data on charm meson $\raa$ and  $\vtwo$ has allowed us to constrain the preferred range of this parameter. It is found that $\raa$ is more sensitive to the average value in the relevant temperature region while $\vtwo$ is more sensitive to the temperature dependence. Taken together, our analysis suggests that a small value $2\pi TD_{s} \simeq (2\sim 4)$ appears to be much preferred in the vicinity of $T_c$ while a relatively strong increase of its value toward higher temperature is favored. The extracted temperature-dependent $2\pi T D_s$ curve is shown in Fig.~\ref{fig:OptimizedCoef} and consistent with other phenomenological analyses as well as lattice calculations.  With the optimized LGR model calculations we've demonstrated a simultaneous description of charm meson $\raa$ and $\vtwo$ observables. We've further made predictions for bottom meson observables in the same model, for which an enhancement of the ratio $\raa(B^{0}_{s})/\raa(B^{+})$ is found in the low to intermediate $\pt$ region with its maximum around $\pt\sim4-6~{\rm GeV}$.  

We end with a discussion on further extending the present analysis for future studies. An important step further is to go beyond the linear ansatz Eq.~\ref{eq:Assum} for the spatial diffusion constant. We plan to further employ a multi-term nonlinear ansatz and to use Bayesian inference for efficiently extracting the full temperature dependence.  Another improvement would be the inclusion of RHIC data in the analysis. Compared with LHC, the RHIC fireball shall be more sensitive to the near-Tc region and would help further constrain the transport coefficient there. It has also been noticed that the major uncertainty in nailing down the temperature dependence lies in the high temperature end, which has not been well constrained by current data. One possibility is to explore the extremely central collisions (e.g. top $1\%$ events of multiplicity) at the highest LHC energies, which should produce a fireball that has more fraction of its space-time evolution in the high temperature region and thus becomes more sensitive to the behavior of QGP in that region. We also plan to explore this idea in the future. 

\begin{acknowledgments} 
The authors are grateful to Miklos Gyulassy, Shuzhe Shi, Hongxi Xing and Pengfei Zhuang for helpful discussions and communications. S.~L. is supported by National Science Foundation of China (NSFC) under Grant Nos.11847014 and 11875178,
China Three Gorges University (CTGU) Contracts No.1910103,
Hubei Province Contracts No.B2018023, 
China Scholarship Council (CSC) Contract No.201807620007,
and the Key Laboratory of Quark and Lepton Physics Contracts No.QLPL2018P01.
J.~L. is supported by the National Science Foundation under Grant No.PHY-1913729. 
The computation of this research was performed on IU's Big Red II cluster,
which was supported in part by Lilly Endowment, Inc., through its support for the
Indiana University Pervasive Technology Institute, and in part by the Indiana METACyt Initiative.
The Indiana METACyt Initiative at IU was also supported in part by Lilly Endowment, Inc.
\end{acknowledgments}
\bibliography{HFT}
\end{document}